\documentstyle[preprint,aps,graphicx]{revtex}

\makeatletter
\def\hlinewd#1{\noalign{\ifnum0=`}\fi
\hrule \@height #1 \futurelet \reserved@a\@xhline}
\def\hwhiteline{\noalign
{\ifnum0=`}\fi\hrule
\@height 0pt\vskip 1.0ex\futurelet \reserved@a\@xhline}
\makeatother
\def\gray{\special{ps: 0.40 setgray}}
\def\black{\special{ps: 0.0 setgray}}
\def\red{
}
\newcommand{\mydraft}{
\newcount\timecount
\newcount\hours \newcount\minutes  \newcount\temp \newcount\pmhours

\hours = \time
\divide\hours by 60
\temp = \hours
\multiply\temp by 60
\minutes = \time
\advance\minutes by -\temp
\def\hour{\the\hours}
\def\minute{\ifnum\minutes<10 0\the\minutes
       \else\the\minutes\fi}
\def\clock{
\ifnum\hours=0 12:\minute\ AM
\else\ifnum\hours<12 \hour:\minute\ AM
\else\ifnum\hours=12 12:\minute\ PM
       \else\ifnum\hours>12
        \pmhours=\hours
        \advance\pmhours by -12
        \the\pmhours:\minute\ PM
        \fi
       \fi
\fi
\fi
}
\def\fullclock{\hour:\minute}
\begin{centering}
\gray
\font\Hugett  =cmtt12 scaled\magstep4
\hbox{\Hugett Draft:\today,\clock}
\black
\end{centering}
\vskip -1.7cm
$\phantom{a}$
} 

\def\beq#1{\begin{equation} \label{#1}}
\def\eeq{\end{equation}}
\def\bra#1{\left\langle #1\right\vert}
\def\ket#1{\left\vert #1\right\rangle}

\newskip\humongous \humongous=0pt plus 1000pt minus 1000pt

\newif\ifdtup

\def\TT{\hbox{\small $\bar{\bf 3}{\bf 3}$}}
\def\SS{\hbox{\small $\bar{\bf 6}{\bf 6}$}}
\def\ss{\hbox{\small $[\bar Q Q]_1\, [X]_1$}}

\def\3s{\hbox{\small $\bar{\bf s}\bar{\bf 3}$}}
\def\6s{\hbox{\small $\bar{\bf s}{\bf 6}$}}
\def\ds{\displaystyle}

\def\QuQd{\hbox{$Qu\bar Q \bar d$}}
\def\RR{\hbox{${\cal R}$}}
\def\mycomm#1{\hfill\break\strut\kern-3em{\red\tt ====> #1\black}\hfill\break}

\def\tallstrut{\vrule height 2.3ex depth 0.0ex width 0pt}
\begin{document}
{\tighten
    \preprint {\vbox{
     \hbox{$\phantom{aaa}$}
     \vskip-0.5cm
\hbox{TAUP 2865/07}
\hbox{WIS/17/07-OCT-DPP}
\hbox{ANL-HEP-PR-07-88}
}}

\title{Possibility of Narrow High-Mass Exotic States}
\author{Marek Karliner\,$^{a}$\thanks{e-mail: \tt marek@proton.tau.ac.il}
\\
and
\\
Harry J. Lipkin\,$^{a,b}$\thanks{e-mail: \tt
ftlipkin@weizmann.ac.il} }
\address{ \vbox{\vskip 0.truecm}
$^a\;$School of Physics and Astronomy \\
Raymond and Beverly Sackler Faculty of Exact Sciences \\
Tel Aviv University, Tel Aviv, Israel\\
\vbox{\vskip 0.0truecm}
$^b\;$Department of Particle Physics \\
Weizmann Institute of Science, Rehovot 76100, Israel \\
and\\
High Energy Physics Division, Argonne National Laboratory \\
Argonne, IL 60439-4815, USA\\
}
\maketitle

\begin{abstract}%
\strut\vskip-1.0cm
Narrow high-mass states can arise 
despite large phase space 
when two nearly degenerate states are
coupled to the same dominant decay mode. Mixing via a final-state interaction
loop diagram can produce one very broad state and one narrow state.
Such a situation is generic in exotic hadrons where a color singlet with 
given flavor and spin quantum numbers can be constructed with
two distinct internal color couplings of quarks. The simplest realization 
of this idea are  the $Q \bar Q q \bar q$ tetraquarks containing two heavy
and two light quarks. We discuss possible experimental implications,
including recent data from Belle.
\end{abstract}

\vfill\eject.

\section{Narrowing of Widths by Mixing}
 
High-mass resonant states containing heavy quark $Q\bar Q$ components are expected to
decay with large
widths into heavy quarkonium $Q\bar Q$ states with one or two additional pions; 
e.g $J/\psi \pi$, $J/\psi \pi\pi$, $\Upsilon \pi$, and $\Upsilon \pi\pi $ 
if phase space is available. 
One example is the case of the
$Qu\bar Q \bar d$ tetraquarks whose masses have been
shown in a number of cases to be comparable to the masses of two separated
mesons\cite{newpentel}. 
Most calculations predict that such states are above the masses of the
separated $Q \bar Q$ and
$u \bar d$ mesons. The $cu\bar c \bar d$ and  $bu\bar b\bar d$ states can 
therefore decay into states like $J/\psi \pi$ and $\Upsilon \pi$ with large
widths. 

Exotic hadrons such as $Qu\bar Q \bar d$ differ from ordinary mesons and
baryons in an important way:
a color singlet with
given flavor and spin quantum numbers can be constructed with
two distinct and nearly degenerate internal color couplings of quarks. 
Therefore in the decay
of such states
there is a possibility that two nearly
degenerate states can be
mixed by a final state rescattering to produce one very broad state and a
comparatively narrow state. 

We follow the approach used \cite{nupenwid} in a similar situation with  a
two-state system coupled to a single dominant decay mode. The mixing via loop
diagrams has been shown to create a decoupling of one of the eigenstates from
this dominant decay mode. Some earlier examples are $\omega-\phi$ 
mixing\cite{katzozi}, the mixing of the strange axial vector
mesons\cite{axial}, and the ``ideal mixing" decoupling the $KN$ 
decay mode in $P$-wave decays of some negative parity strange baryons  
\cite{faiman,karlisg}. 
The suggestion that a narrow width of the putative $\Theta^+$ 
pentaquark might be
due to a decoupling mechanism has also been made in a different 
context\cite{naftali}.

We apply this approach to $Q \bar Q X $ models  where X denotes some system of
quarks and antiquarks and show that a single dominant decay mode can be
decoupled to a good approximation from one of the two diquark-antidiquark
eigenstates. The simplest case is  a tetraquark where $X$ denotes a light
$q\bar q$ pair, but our treatment is more general and includes multiquark
systems with more quarks. 

\vskip1em
\noindent
{\bf Two allowed color couplings}
\vskip1em

Since the heavy quark is a color triplet, 
there are two allowed color couplings
for a $Q \bar Q $: a color singlet and a color octet. There are therefore two
possible couplings for the $Q \bar Q X $ system to a color singlet: the
singlet-singlet and the octet-octet.

It is convenient to define another basis for describing these two states. 
This emphasizes the diquark-antidiquark couplings which appear as mass
eigenstates in tetraquark models~\cite{newpentel}.

\vskip 1em
\noindent
{\bf The \boldmath ($\TT$ -- $\SS$) \unboldmath basis.}
\begin {itemize}
\item The $\ket{\TT}$ state is a color singlet state in which the color triplet 
heavy quark $Q$ is
coupled with a color triplet light quark system to make  a color antitriplet
while the color antitriplet heavy antiquark $\bar Q$ is  coupled with the 
remaining 
light  quark system to make  
color triplet. 
\item The $\ket{\SS}$ state is a color singlet state in which the color triplet 
heavy quark $Q$ is
coupled with a color triplet light quark system to make  a color sextet
while the color antitriplet heavy antiquark $\bar Q$ coupled with the remaining 
light  quark system to make  
color antisextet. 
\end{itemize}
This basis is defined to include a tetraquark where the light quark
system consists of a single antiquark and a single quark. This is a 
complete basis for tetraquark states. For more complicated multiquark 
states there are additional couplings for the light quarks to larger 
color multiplets than the color sextet. We neglect these couplings here. 
Thus our treatment is exact for tetraquark states but may well be a good 
approximation for higher multiplet states. 

\vskip1em
The eigenstates of the tetraquark mass matrix, 
denoted by $\ket {[Tet]_S}$ and 
$\ket{[Tet]_L}$,
by analogy with the kaon eigenstates,  will have the form 
\beq{ThetaSphi}
\begin{array}{cc}
\ds
\ket {[Tet]_S} \equiv 
\cos \theta \cdot \ket {\TT} +
\sin \theta \cdot \ket {\SS} 
\hfill\\
\\
\ket {[Tet]_L} \equiv 
\sin \theta \cdot \ket {\TT} -
\cos \theta \cdot \ket {\SS} 
\end{array}
\end{equation}
where the mixing angle $\theta$ is determined by the 
diagonalization of the mass matrix.
    
 We now consider the case where there is a dominant decay mode to a 
final state of a separated quarkonium color singlet state and a color 
singlet light quark state; e.g. $J/\psi \pi$. We denote this state by 
$\ss$.
 \vskip1em
Since each of the two states (\ref{ThetaSphi})
can decay  to the  $\ket{\ss}$ final 
state, we
define their decay transition matrix elements respectively as
\beq{alphabet}
\bra{\ss} T \ket{\TT} \equiv \alpha; ~ ~ ~  \bra{\ss} T \ket{\SS} \equiv \beta
\end{equation}
We then find that these two states can be mixed by a loop diagram
\beq{loop}
\ket{[Tet]_i} \rightarrow \ss \rightarrow \ket{[Tet]_j}
\end{equation}
as schematically shown in Fig.~1.

\vbox{
\strut
\hfill\break
\hfill\break
\centerline{
\includegraphics[width=10em,clip=true,angle=90]{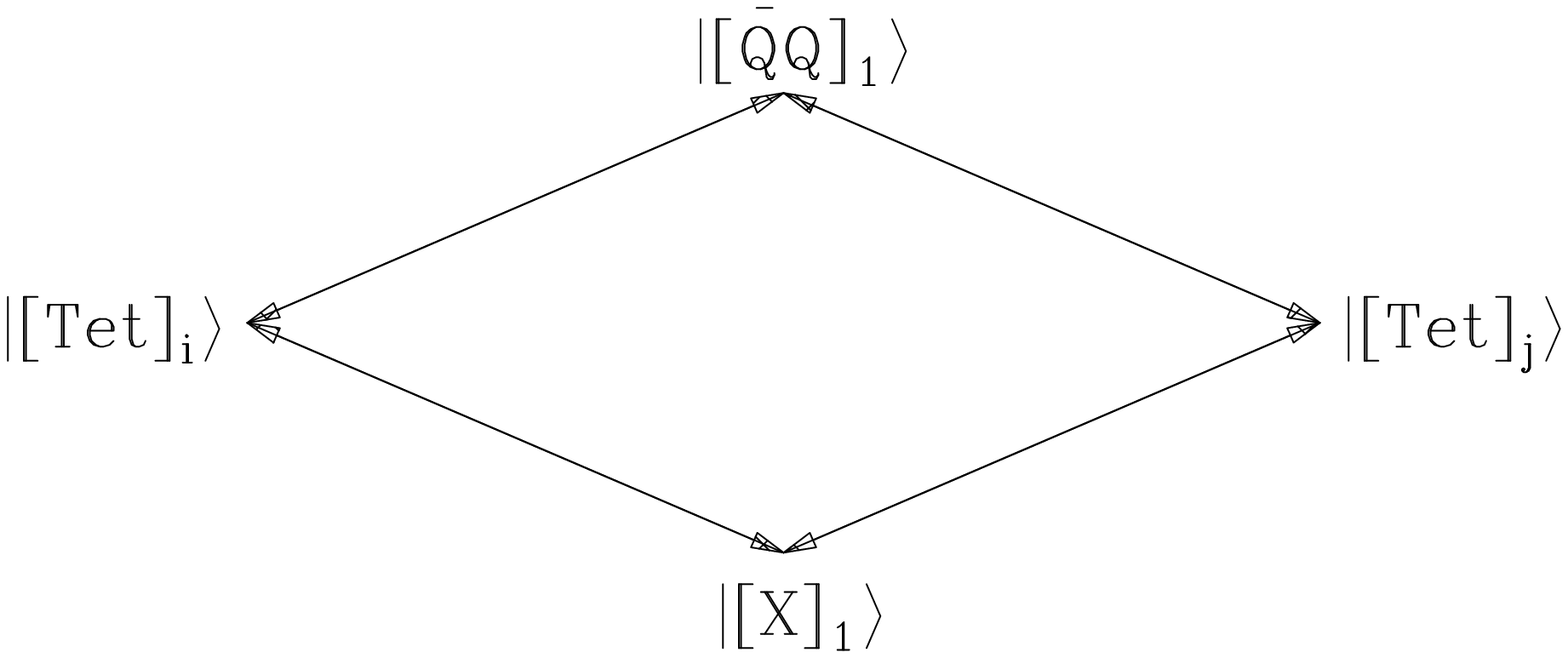}
}
\vskip0.5cm
{\small \em Fig. 1.  
The loop diagram responsible for mixing between the two tetraquark states
$\ket{Tet}_i$ and 
$\ket{Tet}_j$ through a virtual decay to the $\ss$ final state.}
\hfill\break
}

\noindent
where $[Tet]_i$ and $[Tet]_j$ denote any two  states.
The contribution of this loop diagram to the mass matrix is 
\beq{loopmass}
M_{ij} \propto \cdot \bra{[Tet]_i} T \ket {\ss}
\bra{\ss} T \ket {[Tet]_j}
\end{equation}
The matrix $M_{ij}$ is seen to have a determinant of zero. Thus one of the two
eigenstates has the eigenvalue zero and is completely decoupled from the
\ss\ final state. 
In the \hbox{$\TT$ -- $\SS$} basis this contribution to
 mass matrix takes the form
\beq{loopmassA}
M_{ij}  \propto
\pmatrix{
\alpha^2  & \alpha \beta
\cr 
\cr 
\alpha \beta  & \beta^2}
\end{equation}
We first consider the approximation where 
$\ket {\TT}$ and $\ket {\SS}$ are degenerate. 
The mass matrix is dominated by the loop diagram contribution (\ref{loopmass})
and other contributions are neglected. The mass eigenstates
(\ref{ThetaSphi}) are:
\beq{ThetaS}
\ket {[Tet]_S} =
\frac {
\alpha\, \cdot \ket {\TT}
+
\beta\, \cdot \ket {\SS}
}{\sqrt {\alpha^2 + \beta^2} \tallstrut }
\end{equation}
\beq{ThetaL}
\ket {[Tet]_L} = 
\frac {
\beta\, \cdot \ket {\TT}
-
\alpha\, \cdot \ket {\SS}
}{\sqrt {\alpha^2 + \beta^2} \tallstrut }
\end{equation}
 The eigenvalues are $\lambda_S \,\propto\, (\alpha^2 + \beta^2)$ and 
$\lambda_L=0$.
Then  
\beq{decThetaL1} 
\bra{\ss} T \ket {{[Tet]_L}}  \propto \ \beta \cdot \bra{\ss} T \ket
{\TT}  - \alpha \cdot\bra{\ss} T \ket {\SS} = 
\beta \alpha - \alpha \beta
=0
\end{equation}

Thus in this approximation the state ${[Tet]_L}$ is forbidden to decay into the 
$\ss$ final state, while the $[Tet]_S$ should have a normal hadronic width of
hundreds of MeV and probably escape observation against the continuum
background. The decoupling of ${[Tet]_L}$ results from a destructive
interference between the decay amplitudes of the $\ket{\TT}$ and
$\ket{\SS}$ configurations.

The lowest-lying states above the lowest quarkonium pseudoscalar and vector 
states, which we denote by $\eta_Q$ and $\Psi_Q$ are states with one or two
additional pions.  The $\eta_Q\pi^+$ and $\Psi_Q\pi^+$ states both have isospin
1 and respectively even and odd $G$-parity.  Thus any exotic charged state which
has isospin 1 and definite $G$-parity can only decay to one of these two states.
This decay will generally have the largest phase space, be the dominant decay
mode for both states and one decay can be suppressed by mixing. The same is
true for exotic neutral states with isospin  zero, since the isoscalar 
$\eta_Q2\pi$ and $\Psi_Q2\pi$ states respectively have odd and even $G$-parity.         
\section{Estimate of the effects when the two states are not degenerate}

A more precise calculation, not feasible at present, will consider other
contributions to the  mass matrix in addition to the loop diagram.  
However, we can show by a rough calculation how the  width to
the dominant decay mode is reduced by a considerable factor if the ratio of the
mass splitting $\delta m$ between the two nearly degenerate states is small
in comparison with the width $\Gamma$ of the broad state.
 
The mass splitting can be treated as a small perturbation which changes the
state $\ket {[Tet]_L}$ by a small amount
\beq{pert}
\ket {[Tet]^{pert}_L} = \ket {[Tet]_L} + \epsilon \ket {[Tet]_S}
\end{equation}
where $\epsilon$ is a small parameter to be determined by the detailed dynamics.
The transition matrix element to the dominant decay mode is then
\beq{pertrans}
\bra{\ss} T \ket{[Tet]^{pert}_L} = \epsilon \bra{\ss} T  \ket {[Tet]_S}
\end{equation}
The width of the perturbed state $\delta \Gamma$ is proportional to the square
of the transition matrix element. Thus the ratio of this width to the width of
the broad state $\Gamma$ is 
\beq{deltagam}
\frac{\delta \Gamma}{\Gamma} = \frac {[\bra{\ss} T \ket{[Tet]^{pert}_L}]^2}
 {[\bra{\ss} T  \ket {[Tet]_S}]^2} = \epsilon^2 
\end{equation}
In first order perturbation theory, the perturbation of the wave function is
given by the ratio of the perturbation to the mass difference between the states
that are mixed by the perturbation. Here the perturbation is the mass difference
$\delta m$ between the two nearly degenerate states, and the mass difference
between the unperturbed states is the contribution $\Gamma$ from the loop
diagram. Thus  
\beq{epsilon_squared}
\frac{\delta \Gamma}{\Gamma} = \frac {[\bra{\ss} T \ket{[Tet]^{pert}_L}]^2}
 {[\bra{\ss} T  \ket {[Tet]_S}]^2} = \epsilon^2 
 = O\left(\left[ \frac {\delta m}{\Gamma}\right]^2\right) 
\end{equation}

\section{The  \boldmath $Q \lowercase{u} \bar Q
\bar{\lowercase{d}}$  \unboldmath tetraquark}

As an example we consider the two color couplings
of the diquark-antidiquark configuration $Q u \bar Q \bar d$ 
where $Q$ denotes a heavy quark with a mass $m_Q=\xi m_u$.
These are denoted respectively as $\TT\,\QuQd$
and $\SS\,\QuQd$ for the triplet-antitriplet and sextet-antisextet couplings.
In a harmonic
oscillator model where spin is neglected 
the ratio \RR\ of their masses 
has been shown to be\cite{newpentel}
\beq{EDDbuburatN}
\RR \equiv
{{M(\TT\,\QuQd)}\over{M(\SS\,\QuQd)}}
= \frac{\sqrt{6}\cdot 2(\xi + 1) +4\sqrt{\xi}}
{\sqrt{3}\cdot 2(\xi + 1) +2\sqrt{10\,\xi}}
\end{equation}

If we substitute the  values of the constituent quark masses obtained by
fitting the ground state meson and baryon spectra\cite{NewPenta},
$m_u=360$ MeV, $m_c=1710$ MeV, $m_b=5050$ MeV,
we obtain $\RR \approx 1.09$ and $\RR \approx 1.17$ for $c$ and $b$
quarks, respectively.
These are sufficiently close to be serious candidates for mixing.
In general $\RR$ depends only weakly on $\xi$ and
 stays fairly close to 1 for a rather wide range of value of $\xi$, 
as shown in Fig.~2.
In turn, by eq.~(\ref{epsilon_squared}) this implies that the destructive
interference between the decay amplitudes of the $\TT\,\QuQd$ and 
$\SS\,\QuQd$ states will keep the width of the tetraquark narrow.

\vbox{
\strut
\hfill\break
\hfill\break
\centerline{
\includegraphics[width=25em,clip=true,angle=90]{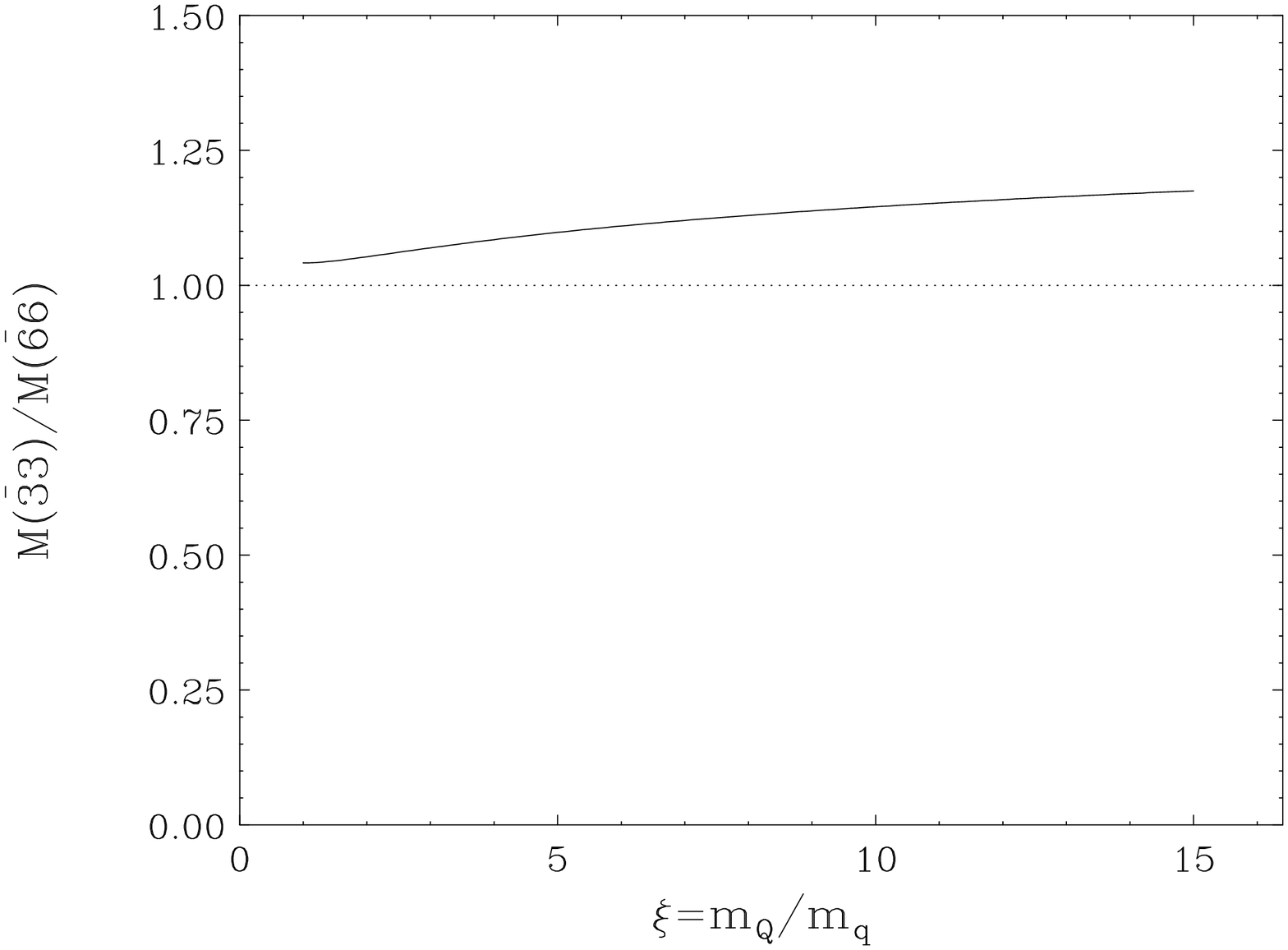}
}
\vskip0.5cm
{\small \em \strut\kern4em Fig. 2.  The ratio 
$\RR \equiv {{M(\TT\,\QuQd)}/{M(\SS\,\QuQd)}}$ as function of 
$\xi \equiv m_Q/m_u$.}
\hfill\break
} 

\section{conclusion}

For states below the threshold for producing charmed or bottom pairs but above
the threshold for producing heavy quarkonium and a pion the mixing mechanism
can produce one state where the amplitude for producing heavy quarkonium and a
pion is suppressed. This is relevant to states like the charmonium and
bottomonium states. This also
answers the point that the tetraquark analysis which does not consider 
mixing\cite{newpentel} has two states which can both decay into heavy quarkonium and a pion and
therefore should both be broad\cite{richard}.

   There is also a mechanism here for suppressing the production of
charmed or bottom pairs. Since these can be produced from either the $\TT$ or
$\SS$ states, there is a possibility that these amplitudes for producing charmed
or bottom pairs can be suppressed by an accidental cancellation. The
relative amplitudes for production from $\TT$ and $\SS$ states are model
dependent and depend upon the parameters $\alpha$ and $\beta$ which depend upon
the tetraquark wave functions. At masses far enough above the threshold
like 700 MeV there will be very many tetraquark configurations, each with
different values of $\alpha$ and $\beta$. Thus the probability that one will
have values close to the value that cancels by accidental cancellation is
not negligible. Only one state in a very rich spectrum of excited
states is needed to do this.

	The recently reported $\psi^\prime \pi^+$ resonance seen by 
Belle\cite{:2007wga} is an example of an exotic state which is not 
observed in what is expected to be its dominant decay mode; namely the 
$J/\psi \pi^+$ decay mode. It is tempting to invoke the mixing mechanism 
for the suppression of this dominant decay mode. However, other 
energetically allowed decay modes like $D^*\bar D$ should also be seen 
unless there are additional accidental cancellations. To investigate 
whether this resonance can be described by the mixing mechanism more 
experimental information is needed beyond its width and the fact that it 
has a $\psi^\prime \pi^+$ decay mode; e.g. spin, parity, other decay modes 
and branching ratios.
 
\section*{Acknowledgements}

The research of M.K. was supported in part by a grant from the
Israel Science Foundation administered by the Israel
Academy of Sciences and Humanities.

%
\catcode`\@=11 
\def\references{
\ifpreprintsty \vskip 10ex
%
\hbox to\hsize{\hss \large \refname \hss }\else
\vskip 24pt \hrule width\hsize \relax \vskip 1.6cm \fi \list
{\@biblabel {\arabic {enumiv}}}
{\labelwidth \WidestRefLabelThusFar \labelsep 4pt \leftmargin \labelwidth
\advance \leftmargin \labelsep \ifdim \baselinestretch pt>1 pt
\parsep 4pt\relax \else \parsep 0pt\relax \fi \itemsep \parsep \usecounter
{enumiv}\let \p@enumiv \@empty \def \theenumiv {\arabic {enumiv}}}
\let \newblock \relax \sloppy
    \clubpenalty 4000\widowpenalty 4000 \sfcode `\.=1000\relax \ifpreprintsty
\else \small \fi}
\catcode`\@=12 

\end{document}